\documentclass[aps,prx,10pt,
               superscriptaddress,
               twocolumn,
               longbibliography,
showpacs]{revtex4-1}

\usepackage[utf8]{inputenc}
\usepackage[english]{babel}  % english language

\usepackage{amssymb,amsmath,amsfonts,amsthm}
\usepackage{mathtools}
\usepackage{verbatim}
\usepackage{enumerate}
\usepackage{graphicx}
\usepackage{hyperref}
\usepackage{bbm}
\usepackage{booktabs}

\usepackage[caption=false]{subfig}

\usepackage{color}
\definecolor{dred}{rgb}{.8,0.2,.2}
\definecolor{ddred}{rgb}{.8,0.5,.5}
\definecolor{dblue}{rgb}{.2,0.2,.8}
\definecolor{dgreen}{rgb}{.2,0.5,.2}
\theoremstyle{plain}
\theoremstyle{definition}

\newcommand{\bra}[1]{\mbox{$\langle #1|$}}
\newcommand{\ket}[1]{\ensuremath{|#1\rangle}}

\newcommand{\hs}{h}

\newcommand{\be}{\begin{equation}}
\newcommand{\ee}{\end{equation}}

\newcommand{\zrot}{\mathcal{R}_z}
\newcommand{\site}[1]{^{[#1]}}  % qubit/site index
\newcommand{\ind}[1]{_{#1}}  % sequence index

\usepackage{array}
\newcolumntype{C}[1]{>{\centering\let\newline\\\arraybackslash\hspace{0pt}}m{#1}}
\begin{document}

%\newcommand{\captionfonts}{\small}
%\makeatletter  % Allow the use of @ in command names
%\long\def\@makecaption#1#2{%
%  \vskip\abovecaptionskip
%  \sbox\@tempboxa{{\captionfonts #1: #2}}%
%  \ifdim \wd\@tempboxa >\hsize
%    {\captionfonts #1: #2\par}
%  \else
%    \hbox to\hsize{\hfil\box\@tempboxa\hfil}%
%  \fi
%  \vskip\belowcaptionskip}
%\makeatother   % Cancel the effect of \makeatletter

%\title{Controlling Quantum Network Dynamics by Time-Reversal Symmetry Breaking}
\title{Chiral Quantum Walks}

\author{Dawei Lu}
\affiliation{Institute for Quantum Computing and Department of Physics and Astronomy,
University of Waterloo, Waterloo N2L 3G1, Ontario, Canada}

\author{Jacob D.~Biamonte}
\email{jacob.biamonte@qubit.org}
\affiliation{ISI Foundation,
Via Alassio 11/c, 10126
Torino, Italy}
\affiliation{Institute for Quantum Computing and Department of Physics and Astronomy,
University of Waterloo, Waterloo N2L 3G1, Ontario, Canada}

\author{Jun Li}
\affiliation{Institute for Quantum Computing and Department of Physics and Astronomy,
University of Waterloo, Waterloo N2L 3G1, Ontario, Canada}

\author{Hang Li}
\affiliation{Institute for Quantum Computing and Department of Physics and Astronomy,
University of Waterloo, Waterloo N2L 3G1, Ontario, Canada}

\author{Tomi H.~Johnson}
\affiliation{Centre for Quantum Technologies, National University of Singapore,
3 Science Drive 2, 117543, Singapore}
\affiliation{Keble College, University of Oxford, Parks Road, Oxford OX1 3PG,
United Kingdom}
\affiliation{Clarendon Laboratory, University of Oxford, Parks Road, Oxford OX1
3PU, United Kingdom}
\affiliation{ISI Foundation, Via Alassio 11/c, 10126
Torino, Italy}

\author{Ville Bergholm}
\affiliation{ISI Foundation,
Via Alassio 11/c, 10126
Torino, Italy}\

\author{Mauro~Faccin}
\affiliation{ISI Foundation,
Via Alassio 11/c, 10126
Torino, Italy}

\author{Zolt\'an Zimbor\'as}
\affiliation{{Dahlem Center for Complex Quantum Systems, Freie Universit\"at  Berlin, 14195 Berlin, Germany}}
\affiliation{Department of Computer Science, University College London,
Gower St, London WC1E 6BT, UK}
\affiliation{Department of Theoretical Physics, University of the Basque Country
UPV/EHU, 48080 Bilbao, Spain}

\author{Raymond Laflamme}
\affiliation{Institute for Quantum Computing and Department of Physics and Astronomy,
University of Waterloo, Waterloo N2L 3G1, Ontario, Canada}
\affiliation{Perimeter Institute for Theoretical Physics, Waterloo, Ontario,
Canada
}

\author{Jonathan Baugh}
\affiliation{Institute for Quantum Computing and Department of Physics and Astronomy,
University of Waterloo, Waterloo N2L 3G1, Ontario, Canada}
\affiliation{Department of Chemistry, University of Waterloo, Waterloo N2L 3G1,
Ontario, Canada
}

\author{Seth Lloyd}
\affiliation{Massachusetts Institute of Technology, Department of Mechanical
Engineering, Cambridge MA 02139 USA}
\affiliation{ISI Foundation,
Via Alassio 11/c, 10126
Torino, Italy}

% alg methods in qm 03.65.Fd
% qm formalism 03.65.Ca
% q systems with finite H space 03.65.Aa

\begin{abstract}
Given its importance to many other areas of physics, from condensed matter physics
to thermodynamics,  time-reversal symmetry has had relatively little influence on
quantum information science.
Here we develop a network-based picture of time-reversal theory, classifying Hamiltonians and
quantum circuits as time-symmetric or not in terms of the elements and geometries of their
underlying networks.
Many of the typical circuits of quantum information
science are found to exhibit time-asymmetry.
Moreover, we show that time-asymmetry in circuits can be controlled using local gates only, and can simulate time-asymmetry in Hamiltonian evolution.
We experimentally implement a fundamental example in which controlled time-reversal asymmetry in a palindromic quantum circuit leads to near-perfect transport.
Our results
pave the way for using time-symmetry breaking to control coherent transport,
and imply that time-asymmetry represents an omnipresent yet
poorly understood effect in quantum information science.
\end{abstract}

%
%  Wigner separated the possible types of symmetries in quantum theory into those
% symmetries that are unitary and those that are antiunitary. Unitary symmetries
% have been well studied whereas antiunitary symmetries and the physical
% implications associated with time-reversal symmetry breaking have had little
% influence on quantum information science. Here we develop a version of
% time-reversal symmetry theory, classifying time-symmetric and time-asymmetric
% Hamiltonians and circuits in terms of their underlying network elements and
% geometric structures. Many of the
% typical quantum Hamiltonians and circuits of quantum information
% science are found to exhibit time-asymmetry.  We experimentally implement
% the most fundamental time-reversal asymmetric process, applying local gates in
% an otherwise time-symmetric circuit to induce time-reversal asymmetry and
% thereby achieve (i) directional biasing in the transition probability between
% basis states, (ii) the enhancement of and (iii) the suppression of these
% transport probabilities. Our results imply that the physical effect of
% time-symmetry breaking plays an essential role in coherent transport and its
% control represents an omnipresent yet essentially untapped resource in quantum
% transport science.

\maketitle

\section{Introduction}
%Quantum walks, as proposed by Feynman~\cite{feynman1965quantum}, model many
%fundamental and naturally occurring processes and are proven to be a universal
%model of quantum computation~\cite{kempe03quantumwalks,childs2009universal,childs2013universal}.
Controlling probability transfer in
quantum systems is a central challenge faced in several emerging
quantum technologies~\cite{jones1998implementation, vandersypen2001experimental,
Lanyon2010towards, zhang2012digital, nv_opt2013}.
Here we develop an approach based on a complex
network theory viewpoint of quantum
systems~\cite{PhysRevX.3.041007,mulken2011ctqw,childs2009universal,Perseguers2010},
in which probability transfer is directed by the controlled breaking of
time-reversal symmetry, creating a so-called {\it chiral quantum
walk}~\cite{Z13,xiang2013transfer,bedkihal2013transfer,manzano2013transfer, Wong15,
cameron2013transfer}.

The practical importance of time-reversal symmetry breaking stems from the fact
that it is equivalent to introducing
biased probability flow into a quantum system. It thus enables directed state transfer
without requiring a
biased (or non-local) distribution in the initial states, or coupling to an
environment~\cite{Z13,plenio2008dephasing, 2012PhRvL.108b0602S, Bose07}.

This work establishes, in terms of the geometry and edge-weights (or gates) of
the underlying graph, conditions on what makes a Hamiltonian (or circuit)
time-asymmetric. As well as allowing us to classify the time-symmetry of well-known Hamiltonians and circuits,
this knowledge allows us to develop active methods to break time symmetry and bias probability transfer, including
using circuits to simulate time-symmetry breaking Hamiltonians.

As a demonstration, the most basic time-asymmetric process is identified and realized
experimentally using room-temperature liquid state nuclear magnetic resonance
(NMR) on a $3$-qubit system.
We show that time-asymmetry and thus biased
probability transport can be
controlled with limited access to the system, namely by using local
$z$-rotations paired with a naturally occurring (or in our case, emulated)
time-symmetric evolution. Through symmetry-breaking we achieve state transfer
probabilities approaching unity.

Since their recent introduction~\cite{Z13}, continuous time chiral quantum walks
have been studied in the context of energy transport in ultracold atoms and
molecules~\cite{xiang2013transfer}, in non-equilibrium
physics~\cite{bedkihal2013transfer,manzano2013transfer}, as a tool for quantum search \cite{Wong15}, and as a method to achieve near
perfect state transfer~\cite{Z13, cameron2013transfer}.  Our theory extends the
contemporary analysis of time-asymmetry such that it can now apply to gate
sequences---of which the existing theory of time-symmetry becomes a special
case---and presents a network classification of the effect. Our experiments
illustrate how active time-reversal symmetry breaking
can be utilized with existing quantum technologies in the presence of limited
control.

\begin{table*}
      \begin{tabular}{C{3.5cm}C{3.5cm}C{3.5cm}C{3.5cm}C{3.5cm}}
        \toprule\\
        & linear chains
        & trees
        & bipartite graphs
        & non-bipartite graphs\\
        &&(possibly with self-edges)
        &  (only even cycles)
        & (some odd cycles)\\
        &
        \begin{minipage}{3cm}
          \includegraphics[width=\textwidth]{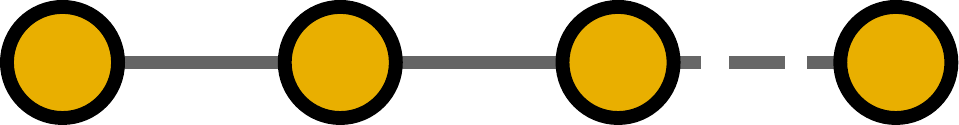}
        \end{minipage}
        &
        \begin{minipage}{3cm}
          \includegraphics[width=\textwidth]{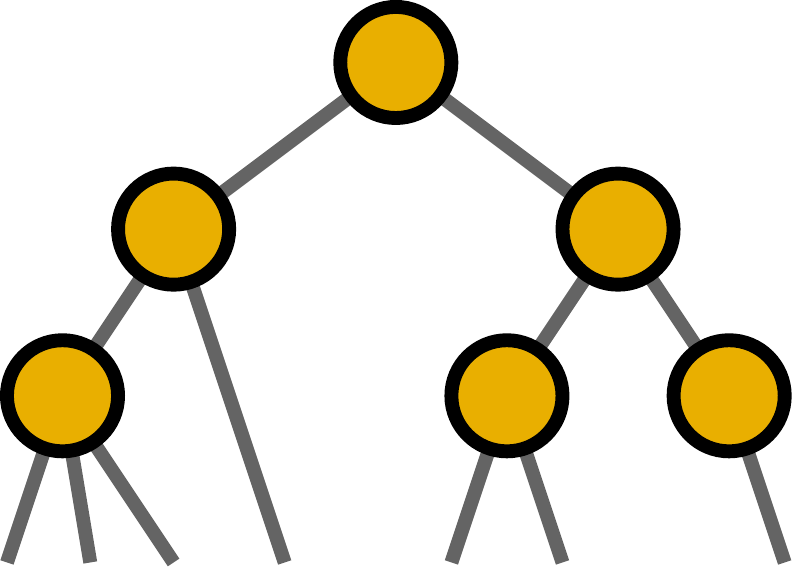}
        \end{minipage}
        &
        \begin{minipage}{3cm}
          \includegraphics[width=\textwidth]{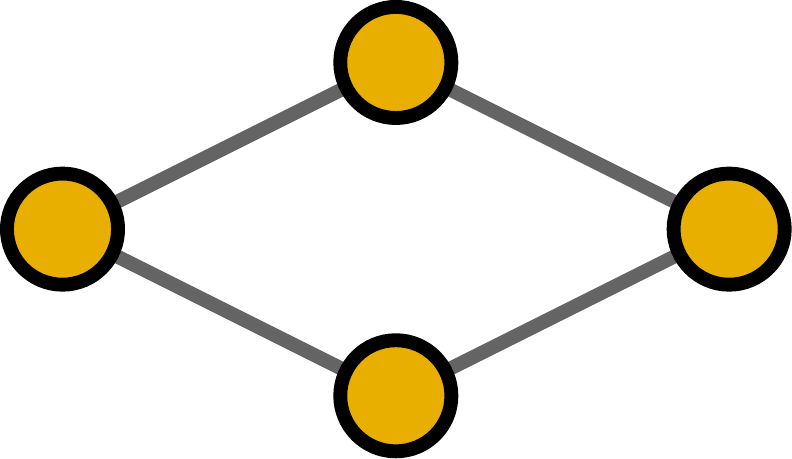}
        \end{minipage}
        &
        \begin{minipage}{3cm}
          \includegraphics[width=0.5\textwidth]{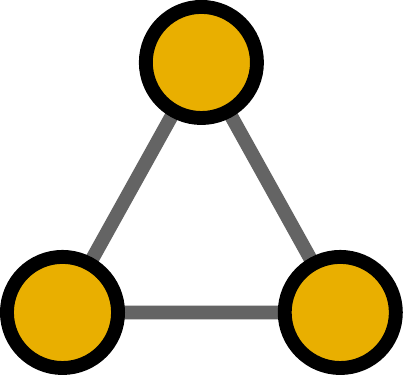}
        \end{minipage}\\[0.5cm]
        \midrule\\
        {\bfseries Probability time symmetric ($\forall\ \alpha_{ij}$)?}
        & \textsc{Yes}
        & \textsc{Yes}
        & \textsc{Yes}
        & \textsc{No}\\
        {\bfseries Probability depends on~$\alpha_{ij}$?}
        & \textsc{No}
        & \textsc{No}
        & \textsc{Yes}
        & \textsc{Yes}\\
        \bottomrule\\
      \end{tabular}
\caption{
In which network geometries do transition probabilities depend on
the complex phases~$\alpha_{ij}$ of the edges of the
(effective Hamiltonian's) internode coupling graph?
%The edges $H_{ij}$ of the graphs are chosen independently as
%$H_{ij} = h_{ij}\mathrm{e}^{\mathrm{i}}$.
We are interested in how the transition probabilities in the site basis
depend on~$\alpha_{ij}$ and if certain values of the~$\alpha_{ij}$ can break time-reversal symmetry.
\label{tab:graph-table}
}
\end{table*}

\section{Chiral quantum walks}
Consider a unitary propagator $U$ acting on a state
space in which we have a preferred basis $\{ \ket{i} \}$. We view~$U$ as performing a quantum walk
over nodes labeled by~$i$, and $U^\dagger$ as performing the time-reversed walk. 
There are two ways in which we consider the walk $U$ to be time-symmetric.

{\it Amplitude time-reversal symmetry} establishes a
relationship $ \bra{i} U \ket{j} = U_{ij} = ( U^\dagger )^\ast_{ij}$
between the inter-node transition amplitudes going forward and backward in time
(where $^\ast$ denotes complex conjugation and $^\dagger$ hermitian
conjugation).  Equivalently, it establishes a
relationship $U_{ij} = U_{ji}$ between opposing transition
amplitudes in the same time direction.

It further implies what we call {\it probability time-reversal symmetry} (PTS):
symmetry of the inter-node transition probabilities going forward and backward in time,
$| U_{ij} |^2 = | (U^\dagger)_{ij} |^2$.
This is equivalent to a lack of directional bias
in the transport between any pair of nodes
in the same time direction:
$| U_{ij} |^2  = | U_{ji} |^2$.

Breaking amplitude time-reversal symmetry is a necessary, but not sufficient, condition
for breaking probability time-reversal symmetry.
In studying internode transport, PTS is the relevant feature.
It leads to a richer classification of processes and is the focus of this work.

\section{Hamiltonian evolutions}
We first consider a unitary propagator $U = e^{-i H t}$ generated by some time-independent Hamiltonian $H$. In our network-based picture, with nodes labeled by~$i$,
we view the matrix elements $\bra{i} H \ket{j} = H_{ij} =h_{ij}e^{i\alpha_{ij}}$ representing Hamiltonian $H$ as forming a complex hermitian adjacency matrix. Here $h_{ij}$ and $\alpha_{ij}$
take only real values. For a non-zero value $H_{ij}$, two nodes $i$ and $j$ are said to be connected by an edge
$e = (i, j)$, and the complex valued edge-weight is given by the value of $H_{ij}$. A special case is an edge that connects $i$ to itself, called a self-edge.
Since it represents the expectation value of the  energy in state $\ket{i}$, the weight $H_{ii}$
of a self-edge must take a real value. Furthermore, self-edges are
only needed if states $\ket{i}$ have different energies $H_{ii}$,
otherwise they may be omitted without changing the underlying physics.
Taken together, the nodes and edges
define a support graph, corresponding to the Hamiltonian~$H$, that we will use
to classify the time-symmetry of~$U$.

We begin by noting that if $\alpha_{ij} = 0$, i.e.\ all edge-weights are real, then both amplitude and probability time-reversal symmetry hold. We will now search for other probability time-symmetric Hamiltonians.

A large class of such Hamiltonians are obtained by considering mapping
$H \mapsto \Lambda^{\dagger} H \Lambda = H'$ (or equivalently
$U \mapsto \Lambda^{\dagger} U \Lambda = U'$), where $\Lambda$ is a diagonal unitary.
Such a mapping will in general affect amplitude time-symmetry. However, it cannot affect transition
probabilities as
$|U'_{ij}|^2 = |\bra{i} e^{-i\Lambda^\dagger H\Lambda t} \ket{j}|^2 =
|\bra{i}\Lambda^\dagger e^{-iH t} \Lambda\ket{j}|^2
= |U_{ij}|^2$,
and hence cannot affect probability time-asymmetry. Thus, we will
call these mappings {\it quasi-gauge symmetry transformations}.

All Hamiltonians that are obtained from a Hamiltonian with real
edge-weights by such gauge transformations are thus probability
time-symmetric.
To give an example, if the graph underlying~$H$ is a tree (of which a linear chain is a special case), there always
exists~\cite{Z13} a~$\Lambda$ which
removes all phases from the edge-weights:
$H_{ij} = h_{ij} e^{i \alpha_{ij}} \mapsto h_{ij}$.
Such Hamiltonians hence never break probability time symmetry. This gives us our first class of time-symmetric evolutions:
those generated by Hamiltonians whose internode couplings
form a tree structure (this may include self-edges).

The class of probability time-symmetric networks is richer than those obtained
by gauge transformations from real Hamiltonians.
To find the other members of this class, we must consider the interference
between walks along different paths. We start by breaking the evolution into
commuting even and odd functions of $H$:
$U = e^{-iHt } = \cosh(iHt) -\sinh(iHt)$.
The probability time-symmetry condition
$|U_{ij}|^2 = |U_{ji}|^2$ can now be expressed as
%\begin{align}\label{eqn:twist} % \frac{\jmath_{mn}(t)}{2}
%\jmath_{ij}(U) /2   &= \sinh(iH)_{ij}\cosh(iH)_{ji} \\
% & -\sinh(iH)_{ji}\cosh(tH)_{ij} ~~~ \forall ~i, j \nonumber
%\end{align}
\begin{equation}\label{eqn:twist}
\sinh(iHt)_{ji}\cosh(iHt)_{ij} = \sinh(iHt)_{ij}\cosh(iHt)_{ji}.
\end{equation}
The physical interpretation of Eq.~\eqref{eqn:twist} is clear:
$\sinh(iHt)_{ij}$ corresponds to transitions along paths of odd length,
and $\cosh(iHt)_{ij}$ to transitions along paths of even length.
Together these terms account for all possible paths between $i$
and~$j$~\footnote{It is not entirely necessary to consider all possible
paths between two nodes as {\em a priori} prescribed by Eq.~\eqref{eqn:twist}.
For a process with $N$ nodes, the Cayley-Hamilton theorem asserts the existence
of a vanishing polynomial with highest degree at most $N$. This implies that
transitions of higher order can be expressed in terms of $N$-long
paths.}. A path from $i$ to itself is called a cycle.
%the weight of each path being inversely proportional to the factorial of its length.

For graph geometries where between each pair of nodes $(i, j)$ there are only
exclusively
even or exclusively odd paths
(equivalent to there being no odd-length cycles in the graph),
probability time-symmetry must always hold as Eq.~\eqref{eqn:twist} is always
satisfied. Such graphs are called
bipartite, where the nodes can be partitioned to two disjoint
sets and non-zero edge-weights
$H_{ij} \neq 0$ only connect nodes of different sets~\footnote{
This is seen by considering a connected graph with
this property, i.e., any two nodes are
connected with paths that have either exclusively of odd or of even lengths.
Let $n$ be an arbitrary node, and denote by $A$ the set of nodes
that are connected to $n$ with paths of even lengths and by $B$ the nodes
that are connected to $n$ with paths of odd lengths.
The union of $A$ and $B$ contain all the nodes, the intersection of
$A$ and $B$ is empty, and any node in $A$ is directly connected only to nodes
in $B$ and vice versa; hence the graph is bipartite.}. Note that this disallows
self-edges, since such edges could be used to make cycles of arbitrary length.
Bipartite graphs can also be understood in terms of gauge transformations, as their structure implies the
existence of a gauge transformation
$H \mapsto \Lambda^{\dagger} H \Lambda = -H$, which immediately implies the probability time-symmetry
condition $|U_{ij}|^2 = | ( U^\dagger )_{ij}|^2$.

This leaves us with two overlapping classes of time-symmetric network
geometries, trees with self-edges and bipartite graphs, where the overlap is the
class of trees without self-edges or, equivalently, bipartite graphs without
simple cycles.
The remaining graph geometries are potentially
time-asymmetric, with the degree of asymmetry determined by the values assigned
to the edge-weights. These findings are summarized with examples in
Table~\ref{tab:graph-table}.

\section{Quantum circuits}
Consider now instead a unitary propagator~$U$ formed by a palindI should be romic circuit
consisting of two-site gates,
$U = \prod\ind{k}^\rightarrow U\ind{k} \prod\ind{k}^\leftarrow U\ind{k}$.
Here the arrows indicate that the second part of the circuit is the reverse of the first,
ensuring the circuit is palindromic, since non-palindromic sequences trivially break time-symmetry and thus do not support control over time-symmetry breaking.
Each circuit is associated with its support graph, which has an edge if and only if
there is a gate in the circuit directly connecting the two sites.

In accordance with the majority of physical implementations, each gate
$U\ind{k} = \exp(-i H\ind{k} t\ind{k} )$ with $H\ind{k} = \sum_{ij \in e_k} (H\ind{k})_{ij} \ket{i} \bra{j}$
acts on two sites connected by the edge~$e_k$.
Like before, we explicitly break apart the two complex off-diagonal elements of each gate Hamiltonian, writing
$(H\ind{k})_{ij} = h_k \exp(i \alpha_k) = (H\ind{k})_{ji}^\ast$, where $e_k = (i,j)$.
We now identify time-symmetric circuits as we did for Hamiltonians.
% \yo{VB: In this case the graph can/should be undirected, I think}

Our starting point is the case in which each gate is amplitude time-symmetric,
$(U\ind{k})_{ij} = (U\ind{k})_{ji}$, or equivalentlyI should be 
$\alpha_k = 0$. This ensures amplitude time-symmetry of the whole palindromic sequence
$U_{ij} = U_{ji}$, and thus also probability
time-symmetry~\footnote{An elementary example of a common circuit family that
induces amplitude symmetric evolutions is palindromic {\it reversible
circuits}.  The analogous case of probability time-reversal symmetry is found
by considering self-inverse circuit families: with
the prototypical example given by {\it Clifford gate} or {\it stabilizer
circuits}. Note that similar palindromic-like circuit descriptions of quantum walks have been investigated in \cite{AsbothObuse13}}. Once again, we find other time-symmetric circuits by considering gauge transformations $U \mapsto \Lambda^{\dagger} U \Lambda$ away from this class of circuits.

Analogous to tree Hamiltonians discussed in the previous section, consider minimal spanning tree circuits, those with only one gate per edge of a tree-like support graph.
There is always a gauge transformation that makes every gate amplitude time-symmetric, $\alpha_k \mapsto 0$, thus implying that $U$ must itself be probability time-symmetric.

Likewise, analogous to bipartite Hamiltonians, any circuit with a bipartite support graph with vanishing diagonal Hamiltonian terms
$(H\ind{k})_{ii} = 0$
is gauge equivalent to its inverse,
and thus naturally exhibits PTS.

\begin{figure*}
    \centering
    \includegraphics[width=\linewidth]{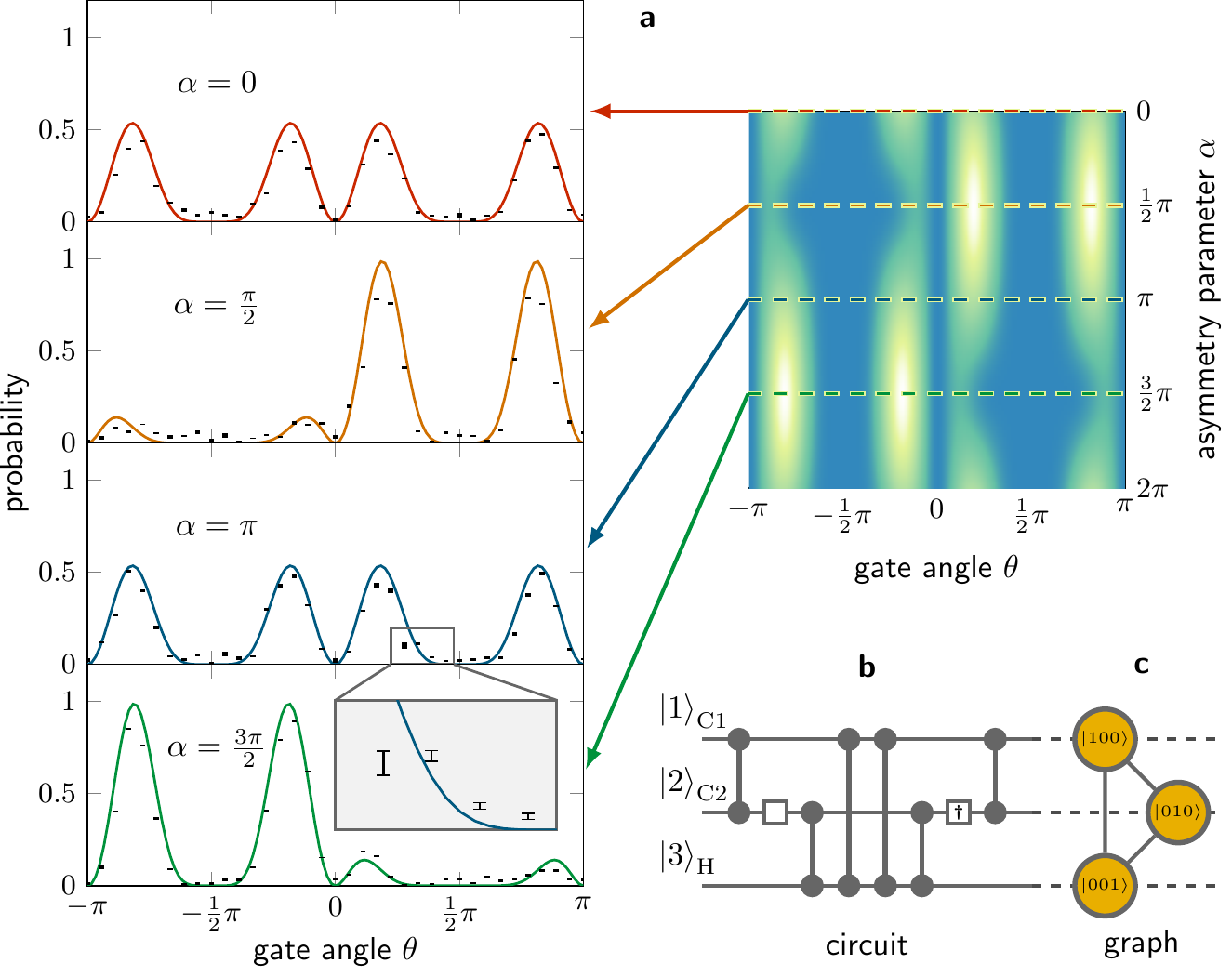}
    \caption{%(color online)
      \textbf{(a)}  The upper right corner depicts the theoretical state-transfer probability
      $|\bra{3} U(\alpha, \theta) \ket{1}|^2$, lighter color indicating higher probability.
      On the left we present four constant-$\alpha$ slices of this function.
      Solid lines are theoretical predictions, dots represent experimental data.
      Dot height represents the experimental error
      (the inset in the bottom plot highlights the error-bars for a few data points).
      The discrepancies between the theoretical and experimental results are explained by
      imperfect radiofrequency control fields and decoherence.
      The rest of the experimental results and a detailed discussion of error
      sources can be found in section~\ref{sec:methods}.
      \textbf{(b)} Quantum circuit diagram corresponding to the experiment.
      All the two-qubit gates are of the form $U\site{ij}(0, \theta)$ defined in
      Eq.~\eqref{eqn:mixing}.
      The left (empty) box in the circuit represents the $\zrot(\alpha)$~gate and
the right box (with dagger $\dagger$) its inverse.
\textbf{(c)}~Graph corresponding to the continuous-time quantum walk (on the single-excitation subspace) simulated by the circuit.
      \label{fig:theo+exp}
    }
\end{figure*}

% Additionally
% with arbitrary diagonal Hamiltonian terms.
% Here the gates act on wires with possible internal tensor structure.
%As long as the palindromic circuit only contains a \emph{two} gates
%corresponding to each graph edge,
%each two-site gate will connect to at least one open wire.
\begin{figure*}[htb]
  \begin{tabular}{cc}
    \begin{minipage}{0.25\textwidth}
      \includegraphics[width=\textwidth]{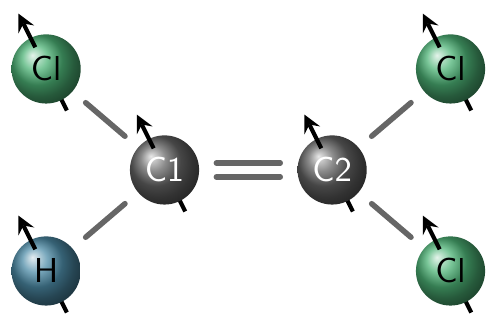}
    \end{minipage}&
    \begin{minipage}{0.6\textwidth}
      \setlength{\tabcolsep}{5pt}
      \begin{tabular}{cccccc}
        \toprule
           (Hz)  & C1      & C2      & H      & $T_1(s)$     & $T_2(s)$\\
        \midrule
        C1       & 21784.6 & -       & -      & $13.0\pm0.3$ & $0.45\pm0.02$\\
        C2       & 103.03  & 20528.0 & -      & $8.9\pm0.3$  & $1.18\pm0.02$\\
        H        & 8.52    & 201.45  & 4546.9 & $8.9\pm0.3$  & $1.7\pm0.2$\\
        \bottomrule
      \end{tabular}
    \end{minipage}\\
    (a)&(b)
  \end{tabular}
\caption{
  \textbf{(a)} Experimental implementation of time-asymmetry controlled transport
in NMR using trichloroethylene in which the two $^{13}$C and one $^{1}$H spins
form a 3-qubit register.
  \textbf{(b)} Hamiltonian parameters for the system.
The diagonal elements are the chemical shifts~$\nu_i$, and the off-diagonal
elements are scalar coupling
strengths~$J_{ij}$.  $T_1$ and $T_2$ respectively are the relaxation and
dephasing time
scales.
\label{fig:molecule}
}
\end{figure*}

%\section{Simulation}
The connection between time-reversal symmetry in circuits and in Hamiltonian evolution can be made even stronger:
%The results for Hamiltonians and circuits are clearly related:
any Hamiltonian evolution can be a simulated by a palindromic circuit of the second-order Trotter type.
For simplicity, consider the Hamiltonian evolution $\exp (-iH t)$ with
$H = \sum_{e \in E} H_e$, where
$H_e = h \sum_{(i,j) = e} \exp(i\alpha_e) \ket{i} \bra{j} +\text{h.c.}$, for some directed edge-set~$E$.
We construct a palindromic circuit~$U$, which for each edge $e \in E$ has a single two-site gate
$U\ind{e} = \exp(-i H\ind{e} t/2)$.
For small enough values of time-parameter $\theta = ht$, the circuit $U \approx e^{-iH t}$ simulates evolution according to~$H$.
Thus palindromic circuits, our focus, naturally include in them the properties of Hamiltonian evolutions.

%{\bf Implementation.}
We consider a simple implementation of the above Trotter-type circuit using a qubit for
each node and working in the single-excitation
subspace. Explicitly, $\ket{j}$ is the state with all qubits in
the state~$\ket{0}$, except $j$, which is in the state~$\ket{1}$. The two-site off-diagonal gates are now two-qubit gates
\begin{equation}\label{eqn:mixing}
U\site{ij}(\alpha, \theta)
= \exp \left(-i \left[ \cos(\alpha)S\site{ij}
+\sin(\alpha ) A\site{ij}\right] \theta/2 \right),
\end{equation}
% NOTE \alpha_{ji} = -\alpha_{ij}, h_{ji} = h_{ij} for consistency
and are generated by combinations of excitation-number-preserving
time-symmetric $S\site{ij} = X\site{i}X\site{j} + Y\site{i}Y\site{j}$ and anti-symmetric $A\site{ij} =
X\site{i}Y\site{j} -Y\site{i}X\site{j}$ Hamiltonians, where $X\site{i}$, $Y\site{i}$ and $Z\site{i}$ are Pauli
matrices acting on qubit $i$.

Many systems naturally possess Hamiltonian terms like
$S\site{ij}$.
Crucially, it is possible to decompose the two-site gate
\begin{align}
\label{eqn:zedeqv}
U\site{ij}(\alpha, \theta)
= \zrot{}\site{j}(\alpha) U\site{ij}(0, \theta) \zrot{}\site{j}{^\dagger}(\alpha),
%e^{-i S\site{ij} \theta/2}
\end{align}
into a symmetric gate generated by $S\site{ij}$ alone, and
local $z$-rotations $\zrot{}\site{j}(\alpha) = e^{-i(\alpha/2)Z\site{j}}$ controlling
time-asymmetry.
Since the $z$-rotations can often be placed in such a way that they combine, often very few rotations are needed to implement $U$.
For example, controlling the probability time-asymmetry in a ring
%where $E = \{ \{1,2 \}, \{2,3 \}, \dots, \{2N,2N+1 \},\{2N+1,1\}  \}$
of $2N+1$ spins requires $2(2N+1)$ symmetric two-site gates but only $1$~pair of
$z$-rotations.

We will now demonstrate how state transfer can be directed by time-symmetry breaking.
The simplest circuit that allows time-symmetry breaking is
\begin{equation}
\label{eq:circuit}
U(3 \alpha, \theta) = U\site{12} U\site{23} U\site{31} U\site{31} U\site{23} U\site{12},
\end{equation}
involving three nodes,
where all the two-site gates have the same parameters: $U\site{ij}(\alpha, \theta)$.
For small enough gate angles~$\theta\,{=}\,ht$ it simulates the evolution $U = e^{-iHt}$ according to the most fundamental Hamiltonian
$H = \hs e^{i \alpha} \left( \ket{1}\bra{2} + \ket{2}\bra{3} + \ket{3}\bra{1} \right) + \mathrm{h.c.}$ that
allows time-symmetry breaking (see Fig.~\ref{fig:theo+exp}c), but we need not be constrained to the regime of small~$\theta$.

\section{Experiment}
\label{sec:methods}

We implement the circuit in Eq.~\eqref{eq:circuit} using NMR techniques in a three-qubit system consisting of
$^{13}$C-labeled trichloroethylene dissolved in deuterated chloroform.
%, as explained in Sec.~\ref{sec:methods}.
The implementation requires six symmetric two-site gates and a pair of $z$-rotations, shown in Fig.~\ref{fig:theo+exp}b.
We directly measure the transition
probabilities $|\bra{i} U(\alpha, \theta) \ket{j}|^2$,
while varying both the gate angle $-\pi< \theta \leq \pi$ and the
time-asymmetry parameter $0 \leq \alpha < 2 \pi$,
where time-symmetry occurs only for $\alpha = n \pi$.

We display our experimental results
for the transport probability from site~1 to site~3
for four values of $\alpha$ and
$\theta= n \pi/18$ in Fig.~\ref{fig:theo+exp}a.  The slice $\alpha = 0$
corresponds to the amplitude and probability
time-symmetric case. The slice $\alpha = \pi/2$ corresponds to maximum probability time-asymmetry.
The slices corresponding to $\alpha=\pi$ and $\alpha=3\pi/2$ represent
a reflection in time $\theta$ of the first two cases.

For the time-symmetric case ($\alpha = 0$) the probabilities of
transporting the excitation to the other two spins are always bounded
from above by~$0.6$.
However, time-asymmetry ($\alpha \neq 0$) allows us to break this
barrier, with transition probabilities approaching unity at
the point of maximal time-asymmetry ($\alpha = \pi/2$),
as shown in Fig.~\ref{fig:theo+exp}a.
%At this maximal time-asymmetry point, the circuit propagator~$U$
%is gauge equivalent to a real matrix.

\begin{comment}
%{\bf Experimental results.}
In the time-symmetric case, we
cannot break time-symmetry because of the palindromic circuit structure in
Fig.~\ref{fig:theo+exp}b.
% nor enhance the transport probabilities.
The experimental data
depicted in the first column of Fig.~\ref{fig:exp} (corresponding to
$\alpha=0,\pi$) illustrates a
good match between theory and experiment.
The top left panel shows that the probabilities of finding the
excitation at spin~2 or spin~3 are always below~$0.6$.
In the third panel of the first column, since time symmetry
cannot be broken under the time-symmetric Hamiltonian, we obtained essentially
the same plot as the left one.

The same gate sequence with $\alpha=\pi/2, 3\pi/2$ results in a large
enhancement of the transport probability.
On the second and fourth panels of Fig.~\ref{fig:exp},
we see that the probabilities of encountering
the excitation in spin~2 or spin~3 approach unity, seen in both
theory and experiment.  Moreover, time-symmetry is again broken by applying
the time-reversed evolution.
\end{comment}

\begin{figure*}[htb]
  \centering
  \includegraphics[width=0.9\textwidth]{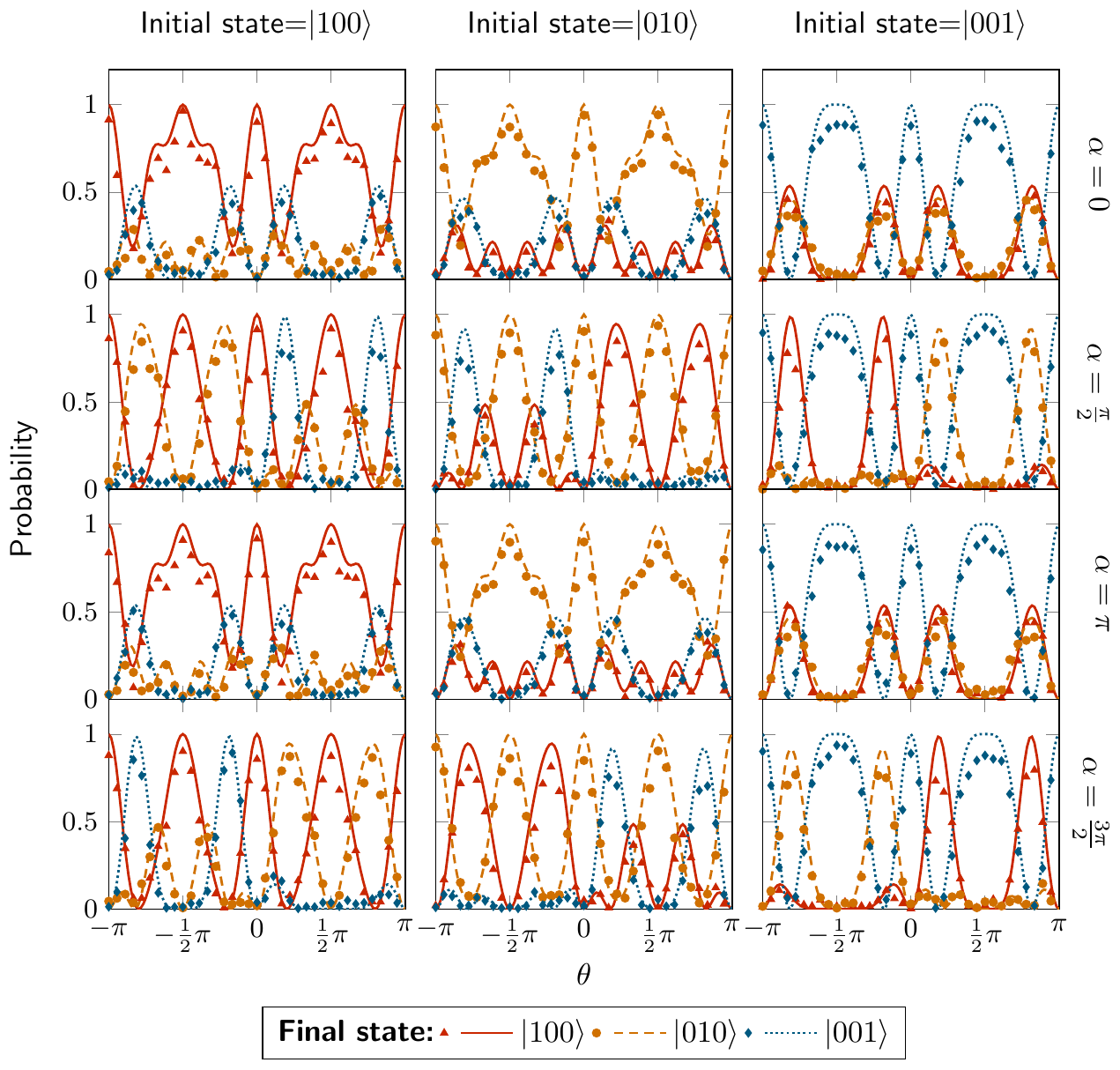}
  \caption{%(color online).
    Experimental results of time-asymmetry controlled transport on the 3-qubit
NMR system.
The three columns correspond to different initial states
($\ket{100}, \ket{010}$ and $\ket{001}$).
The red solid, yellow dashed and blue dotted curves are the
theoretical probabilities of measuring $\ket{100}$, $\ket{010}$ and $\ket{001}$,
respectively.
The triangles, circles and diamonds represent the corresponding experimental results.
Experimental values are measured at 36 equally spaced ($\pi/18$) time steps
in the range from $-\pi$ to $\pi$.
%All probabilities are measured every $\pi/18$ time in experiment.
The plots with $\alpha=0,\pi$ correspond to time-symmetric
gates and its time-reversed evolution (which cannot break time
symmetry). The plots with $\alpha=\pi/2,3\pi/2$ correspond to
time-asymmetric gates and its time-reversed evolution,
which do exhibit time-reversal asymmetry.
}\label{fig:exp}
\end{figure*}

%Besides the experiments with the initial excitation being on spin~1, w
For completeness, we also investigated the cases with the initial excitation localized at
spins 2 and~3. The full results are presented in Fig.~\ref{fig:exp}
--- illustrating similar properties of time-symmetry breaking and
suppression or enhancement of transport probabilities.

The average error of the experimental data relative to the theoretical
predictions is about 6.0\%, and it can be attributed to two main factors:
decoherence and imperfection of GRAPE pulses. The decoherence mainly
originates from $T_2$ relaxation, which induces about 1.5\% signal loss. The
remaining 4.5\% error mostly comes from the imperfection of GRAPE pulses,
as well as a minor inhomogeneity of static and
radio-frequency magnetic fields.

\subsection*{Experimental setup}

All experiments are carried out
on a Bruker~DRX 700~MHZ NMR spectrometer at room
temperature. The sample is  $^{13}$C-labeled trichloroethylene (TCE)
dissolved in deuterated chloroform.
The structure of the molecule is shown in
Fig.~\ref{fig:molecule}a, where we denote C1 as qubit 1, C2 as qubit 2, and H as qubit 3.
The natural Hamiltonian of this system is
\begin{eqnarray}\label{Hamiltonian}
H = &&\sum\limits_{j=1}^3 {\pi \nu_j } Z\site{j}  +
\frac{\pi}{2}(J_{13}Z\site{1} Z\site{3}+J_{23}Z\site{2} Z\site{3})
\nonumber\\
&&+ \frac{\pi}{2}J_{12} (X\site{1} X\site{2} +Y\site{1} Y\site{2} +Z\site{1} Z\site{2}),
\end{eqnarray}
where $\nu_j$ is the chemical shift of the $j$th spin and $J_{ij}$ is the scalar
coupling strength between spins $i$ and $j$. As the difference in the chemical shifts
between C1 and C2 is not large enough to adopt the weak $J$-coupling
approximation~\cite{nmrreview}, these two carbon spins are treated as strongly coupled.
The parameters of the Hamiltonian are determined by iteratively fitting the simulated and experimental spectra,
and presented in the table in Fig.~\ref{fig:molecule}b.

%\begin{figure*}[h] \centering
%\includegraphics[width=0.9\textwidth]{molecule}
%\caption{(color online). (a) Experimental implementation of chiral quantum walk
%in NMR using trichloroethylene in which the two $^{13}$C and one $^{1}$H spins
%form a 3-qubit sample. In the parameter table  the diagonal elements are the
%chemical shifts (Hz), and the off-diagonal elements are scalar coupling
%strengths (Hz).  T$_1$ and T$_2$ are the relaxation and dephasing time scales.
%(b) Quantum circuit of implementing chiral quantum walk. The six time evolutions
%which form a palindromic sequence are dominated by time-symmetric Hamiltonian
%XX+YY or time-asymmetric Hamiltonian XY-YX
%\label{fig:molecule}
%}
%\end{figure*}

Without loss of generality, we will describe the
experimental procedure with spin 1 initially excited, i.e.,  $\ket{100}$ as
the initial state.
Each experiment consists of three main parts:
(A) State initialization: Preparing the system in the pseudo-pure state $\ket{000}$,
and then exciting one spin to the state~$\ket{100}$;
(B) Evolution: Driving the system
%evolved under the symmetric or asymmetric Hamiltonian, in a
through a palindromic quantum circuit;
(C) Measurement: Measuring the probabilities of finding the excitation
at each of the spins.

\paragraph*{(A) State initialization.}
Starting from thermal equilibrium, we first
create the pseudo-pure state
\begin{equation}
\rho_{000}=(1-\epsilon)\openone/8+\epsilon \ket{000} \bra{000},
\end{equation}
using the spatial averaging technique~\cite{spatial}. Here $\epsilon \approx
10^{-5}$ quantifies the polarization of the system and $\openone$ is the $8\times
8$ identity matrix. Next, we apply a $\pi$ pulse on spin 1 to rotate it to the
excited state~$\ket{1}$.
This $\pi$ rotation is realized by a $2$~ms and over
$99.5$\% simulated fidelity GRadient Ascent Pulse Engineering (GRAPE)
pulse~\cite{grape1,grape2}. All GRAPE pulses in the experiment are designed
to be robust against the inhomogeneity of radio-frequency pulses.

\paragraph*{(B) Evolution.}
The initial state will be evolved under
four types of effective Hamiltonians: a symmetric Hamiltonian
$S\site{ij}$ and its time reversed version $-S\site{ij}$,
and the asymmetric Hamiltonian
$A\site{ij}$ and its time reversed version, all obtained from
$S\site{ij}$ using local $\zrot{}$~pulses as shown in Eq.~\eqref{eqn:zedeqv}.
The circuit of the entire sequence is depicted
in Fig.~\ref{fig:theo+exp}b, where
the six two-body interactions form an palindromic circuit for this
3-qubit system.  (Note that the two central gates can be merged into a
single gate, corresponding to a five-gate palindrome.)  We further note that
the gate from Eq.~\eqref{eqn:zedeqv} can be expressed analytically as
\begin{align}
U\site{ij}(\alpha, \theta)
=& \zrot{}\site{j}(\alpha) U\site{ij}(0, \theta) \zrot{}\site{j}{^\dagger}(\alpha) \\
=& \frac{1}{2} \big(1 + Z\site{i} Z\site{j} +\cos(\theta) (1- Z\site{i} Z\site{j}) \nonumber \\
& -i \sin(\theta) \left(\cos(\alpha) S\site{ij} + \sin(\alpha)A\site{ij}\right)\big).
\nonumber
\end{align}

%Each interaction can be represented by a unitary
%operator $e^{-iHt}$, where $H=\pm(X_iX_j+Y_iY_j)$
%in the symmetric case and $H=\pm(X_iY_j-Y_iX_j)$ in the asymmetric case,
%applied for a given time~$t$.

The experiment utilized GRAPE pulses with different lengths
to implement all of the two-body interactions, depending on the $J$-coupling
strength (see the table of Fig.~\ref{fig:molecule}b) of the related two spins. The
three typical lengths of GRAPE pulses for implementing the two-body interactions
are $3$~ms for $J_{12}$, $2$~ms for $J_{23}$ and $8$~ms for $J_{13}$, respectively.
Therefore, the overall run-time of the circuit comprised of all six evolutions is $26$~ms,
which is much less than the decoherence time as seen in
Fig.~\ref{fig:molecule}b. For all four of the Hamiltonian types,
we implemented the circuit 37  times as $\theta$ was chosen to realize
every $\pi/18$ step in $[-\pi, \pi]$. The total number of GRAPE pulses
is 444 and all pulses have simulated fidelities over 99\%.

\paragraph*{(C) Measurement.}
After implementing the circuit, we measure the
probabilities of finding the excitation at each spin, i.e.~the probabilities
of the $\ket{100}$, $\ket{010}$ and $\ket{001}$ states, which
corresponds to standard population measurement in the NMR setup. We use a
$\pi/2$ pulse to rotate spin 2 to the transverse $x-y$ plane and compare the
relative intensities of the transitions with the initial state. Then all three
probabilities can be obtained.

\section{Discussion}
%directing of transport and the breaking of probability time-symmetry,
%whereby the breaking of symmetry to allow directional biasing of
%transition probabilities also supports a greater enhancement of these
%probabilities. We have classified Hamiltonian evolutions and circuits
%that break probability time-symmetry, finding a dichotomy between
%commonly used processes: Hamiltonians and circuits
%with bipartite geometry (e.g.~square networks with no self-edges)
%never break probability time-symmetry, while those with non-bipartite geometry
%(containing odd cycles, e.g.~triangular graphs) can exhibit a high degree of
%symmetry-breaking.
%
%

The behavior of the fundamental laws of physics under time-reversal  has long
remained central to the foundations of
physics~\cite{wigner59,2013PhRvL.111b0504S} and has found use in condensed
matter theory~\cite{peierls1993,hofstadter1976energy,sarma2008hall,
hasan2010topological,dalibard2011artificial}.
By going beyond Hamiltonian-generated quantum walks, considering quantum circuits
and analyzing their time-reversal properties, we obtain
a far richer set of behaviors.
%By going beyond time-independent Hamiltonians to consider quantum circuits
%and analyze their time-reversal properties, we obtain
%a far richer set of behaviors.
This also provides us a set of new tools for controlling transport in quantum systems.

Focusing on the simplest three-qubit circuit that allows time-symmetry breaking,
we experimentally demonstrate
that this asymmetry can lead to noticeably enhanced transition probabilities.
Further, we
show that the circuit's time-symmetry or
lack thereof can be completely controlled by local $z$-rotations.
This is reminiscent of how
one can change the sign of a Hamiltonian by the application
of local gates, which is a common tool in NMR experiments~\cite{nmrreview}.
The most elementary example is reversing the sign of a $\sigma_z$ Hamiltonian with
a $\pi$-pulse about the $x$-axis.
A more interesting example is the so-called magic echo for reversing the sign of the homonuclear
dipolar Hamiltonian~\cite{Rhim71, cho2005multispin}.
%In fact, this is reminiscent of
%our circuit (Fig.~\ref{fig:theo+exp}b) applied twice in succession, first
%with~$\alpha$ and then with $\alpha + \pi$, evolving
%for equal times each---the result is the identity propagator.

We emphasize that time-asymmetric site-to-site transport can only take place in circuits whose circuit graphs have odd cycles (see Table~\ref{tab:graph-table}).
We expect our method of directing quantum transport also to be
applicable in much larger networks.
The amount of enhancement achievable  will most likely be heavily dependent
on the topology of the network and the structure of the gate sequence,
  an interesting topic for future research.\\

\begin{acknowledgments}
VB, JDB, MF, THJ and ZZ completed parts of this study
while visiting the Institute for Quantum Computing at the University of Waterloo,
and acknowledge financial support from the Fondazione Compagnia di San Paolo
through the Q-ARACNE project.  JDB acknowledges the Foundational Questions
Institute (FQXi, under grant FQXi-RFP3-1322) for financial support. THJ
acknowledges the National Research Foundation and the Ministry of Education of
Singapore for funding. ZZ acknowledges support from the EPSRC and by the EU
through the ERC Starting Grant GEDENTQOPT and the CHIST-ERA QUASAR project. DL, JL, HL,
RL  and JB acknowledge Industry Canada, NSERC and CIFAR for financial support.
SL acknowledges the ARO, DARPA, AFOSR, Eni via MITEI, and
Jeffrey Epstein for financial support.
\end{acknowledgments}

\end{document}